\documentstyle[aps,epsf,graphics,multicol]{revtex}

\newcommand{\xy}{\textsl{XY} }
\begin{document}
\draft
\preprint{01-020}
\title{\textsl{XY} model in small-world networks}
\author {Beom Jun Kim,$^1$ H. Hong,$^2$ Petter Holme,$^1$ Gun Sang Jeon,$^3$ 
Petter Minnhagen,$^1$ and M.Y. Choi$^2$}
\address {
$^1$Department of Theoretical Physics, Ume{\aa} University, 
901 87 Ume{\aa}, Sweden \\
$^2$Department of Physics, Seoul National University, Seoul 151-747, Korea \\
$^3$Center for Strongly Correlated Materials Research, Seoul National
University, Seoul 151-747, Korea}

\maketitle

\begin{abstract}
The phase transition in the \xy model on one-dimensional
small-world networks is investigated by means of Monte-Carlo simulations. 
It is found that long-range order is present at finite temperatures, 
even for very small values of the rewiring probability, 
suggesting a finite-temperature transition for any nonzero rewiring probability. 
Nature of the phase transition is discussed in comparison with
the globally-coupled \xy model.
\end{abstract}

\pacs{PACS numbers: 64.60.Cn, 64.60.Fr, 84.35.+i, 74.50.+r}

\begin{multicols}{2}
Examples of complex networks are abundant in many disciplines of science and
have recently attracted much attention~\cite{network}.
Among interesting phenomena is the so called {\em small-world effect}
that a randomly chosen pair of nodes can be connected by
a remarkably small number of intervening nodes.
This effect, first noticed by Watts and Strogatz (WS)~\cite{ws}, 
can be observed in a variety of
real network systems such as the World-Wide Web, 
social networks, and scientific collaboration networks~\cite{www,newman}.
The WS model is based on
a locally highly connected regular network, in which some of the links 
are randomly ``rewired'', creating long-range ``shortcuts''.
In such a model network the small-world effect is usually measured
by the scaling behavior
of the characteristic path length $\ell$, defined to be the average of 
the shortest distance between two nodes: $\ell\sim\log N$ with the network size 
(i.e., the number of nodes) $N$. 
Noteworthily, the small-world phenomena in the WS model
emerge even at very small values of the rewiring probability 
$P \sim {\cal O}(N^{-1})$, which implies that the global feature of the network 
is altered dramatically in the presence of only a tiny fraction of shortcuts.

From the point of view of statistical physics, the above small-world effect
can imply the emergence of global coherence in the presence of shortcuts:
In the absence of shortcuts, global coherence is difficult to achieve
since the information to make each element have the same state should travel
long distance of the order of the network size. 
As the number of shortcuts is increased, on the other hand, 
long-range connections become available, assisting the system to behave 
as a whole. 
The significance of such a shorter global length scale 
has been tested with many statistical physical problems 
studied on a small-world network topology. 
Included are the signal propagation speed~\cite{ws},
synchronizability~\cite{ws}, dynamics of Hodgkin-Huxley neurons~\cite{FHCS},
epidemiology~\cite{KA,MN}, percolation~\cite{MN,NW,percol},
and---most relevant to the present study---the Ising model, 
where a nonvanishing order parameter has been demonstrated
in the presence of a vanishingly small fraction of shortcuts~\cite{ising}.

The \xy model, describing two-dimensional (2D) spin (i.e., planar rotor) 
systems or superconductors and superfluids,
is one of the most well-known systems in statistical physics.
In one and two spatial dimensions, the Mermin-Wagner theorem predicts that 
the \xy model with only local interactions should not possess long-range 
order at any nonzero temperatures.  Nevertheless, in two dimensions, it has been
shown that although true long-range order is not possible, 
quasi-long-range order appears at finite temperatures~\cite{BKT}. 
As the \xy model includes more and more long-range interactions, 
one expects that the system can display true long-range order.
In this work we study the \xy model on the WS model of small-world networks
and investigate the possibility of long-range order
as the number of shortcuts or long-range interactions is increased. 

The Hamiltonian for the \xy model reads
\begin{equation}
H = - \frac{1}{2}\sum_{i\neq j} J_{ij} \cos(\phi_i - \phi_j) ,
\end{equation}
where $\phi_i$ is the angular direction of the 2D spin or 
the phase of the superconducting order parameter at node $i$.
The coupling matrix $J_{ij}$ is given by
\begin{equation} \label{eq:Jij}
J_{ij}  = J_{ji} \equiv \left\{
\begin{array}{ll}
J, & \mbox{if $i$ and $j$ are connected,} \\
0, & \mbox{otherwise.}
\end{array}
\right.
\end{equation}
For example, in the standard 2D \xy model on a square lattice,
we have $J_{ij} = 0$ except for the
four nearest neighbors of site $i$. 
In the WS model for the small-world network~\cite{ws}, the
regular network with $N$ sites is first constructed by connecting
each site to its $2k$ nearest neighbors. 
Each local link is visited once, and
then with the rewiring probability $P$ it is removed and reconnected
to a randomly chosen site. 
In this manner, a small-world network with size $N$ 
is constructed for given $k$ and $P$.
We then study the phase transition
of the \xy model on this small-world network by measuring the order 
parameter
\begin{equation} \label{eq:m}
\langle m \rangle \equiv \left\langle \left|  
                   \frac{1}{N}\sum_j e^{i\phi_j} \right| \right\rangle ,
\end{equation}
where $\langle \cdots \rangle$ denotes the thermal average and the average
over different network realizations, denoted by $[\cdots]$, is also to be taken.
When $P=0$, the system behaves as the one-dimensional (1D) \xy model, 
displaying no long-range order at finite temperatures.  
For $P \neq 0$, the system has many 
long-range interactions in the thermodynamic limit and 
the Mermin-Wagner theorem cannot disprove the
existence of long-range order.
In particular, for $P=1$, it is on a random network with the number of randomly
connected links being $kN$. 

We perform extensive Monte-Carlo (MC) simulations with the standard Metropolis local
update algorithm, to measure various thermodynamic quantities
and moments of the order parameter.
For a  thorough analysis of the MC results, 
we  detect the phase transition in three different methods:
First, Binder's fourth-order cumulant~\cite{binder}
\begin{equation} \label{eq:Binder}
U_N(T) \equiv 1 - \frac{ [\langle m^4 \rangle] } { 3 [\langle m^2 \rangle]^2 }
\end{equation}
for different sizes $N$ should have a unique crossing point at the 
transition temperature $T_c$.
Expanding the cumulant near $T_c$, we write
\begin{equation} \label{eq:Bexpand}
U_N(T) \approx U^* + U_1\left(1-\frac{T}{T_c}\right)N^{1/{\bar\nu}} 
\end{equation}
or
\begin{equation} \label{eq:Bexpand1}
\Delta U_N \equiv U_N(T_1) - U_N(T_2) \propto N^{1/\bar\nu},
\end{equation}
with $T_1$ and $T_2 \,(>T_1)$ picked near $T_c$.
From the above expression,
the critical exponent $\bar\nu$, describing
the divergence of the correlation volume $\xi_V$ at $T_c$~\cite{botet,foot:nubar}
\begin{equation} 
\xi_V  \sim |T - T_c|^{-\bar\nu},
\end{equation}
can be determined.
Our second method is based on the finite-size scaling 
of the order parameter, which exhibits the critical behavior 
$[\langle m \rangle ] \sim (T_c - T)^\beta$
in the thermodynamic limit. 
In a finite-sized system, we expect the behavior
to $[\langle m \rangle ]  = (T_c - T)^\beta f(T,N)$ 
with a function $f$ of at most two arguments $T$ and $N$.
We then use the standard finite-size scaling 
idea that the ratio of the correlation volume to the system size gives the 
argument of the scaling function, and obtain the finite-size scaling form
\begin{equation} \label{eq:mscale}
[\langle m\rangle] = N^{-\beta/\bar\nu} g\left( (T{-}T_c) N^{1/{\bar\nu}}\right) ,
\end{equation}
which leads to a unique crossing point at $T_c$
in the plot of $[\langle m\rangle] N^{\beta/\bar\nu}$ versus $T$.
As discussed later in the present work, the phase transition on the small-world
network is of the mean-field nature, which provides us the third method of locating
the transition temperature: 
With the mean-field value of the critical exponent $\alpha =0$,
the specific heat neither diverges nor vanishes, 
remaining finite possibly with jump discontinuity.
Accordingly, it is plausible to write 
the finite-size scaling form of the specific heat as
\begin{equation} \label{eq:Cvscale}
C_v = h\Bigl( (T{-}T_c)N^{1/{\bar\nu}}\Bigr), 
\end{equation}
which again crosses at $T_c$. 
In the same way as in Eq.~(\ref{eq:Bexpand}),
the expansion of $C_v$ near $T_c$ can be used to determine $\bar\nu$.

All the above finite-size scaling forms are based on the assumption
that we have only two spatial scales in the system: the size $N$
and the correlation volume $\xi_V$ diverging at $T_c$.  
In fact it is known that the small-world network 
has an additional length scale,
the typical distance between the ends of shortcuts, 
given by $\zeta = (kP)^{-1}$~\cite{NW}.  
Accordingly, in the presence of the three competing scales $(N, \xi, \zeta)$, 
the standard finite-size scaling function 
should take the form $\chi(\xi/N, \zeta/N)$.
Here we focus on sufficiently large systems with $N$ much larger than $\zeta$, 
where $\chi(\xi/N, \zeta/N)$ may be approximated as $\chi(\xi/N, 0)$.
This leads to the above-mentioned finite-size scaling forms without $\zeta$
[Eqs.~(\ref{eq:Bexpand}), (\ref{eq:mscale}), and (\ref{eq:Cvscale})], 
which become more precise as the system size grows.
This is manifested below when the MC results are presented.

\begin{figure}
\centering{\resizebox*{!}{6.0cm}{\includegraphics{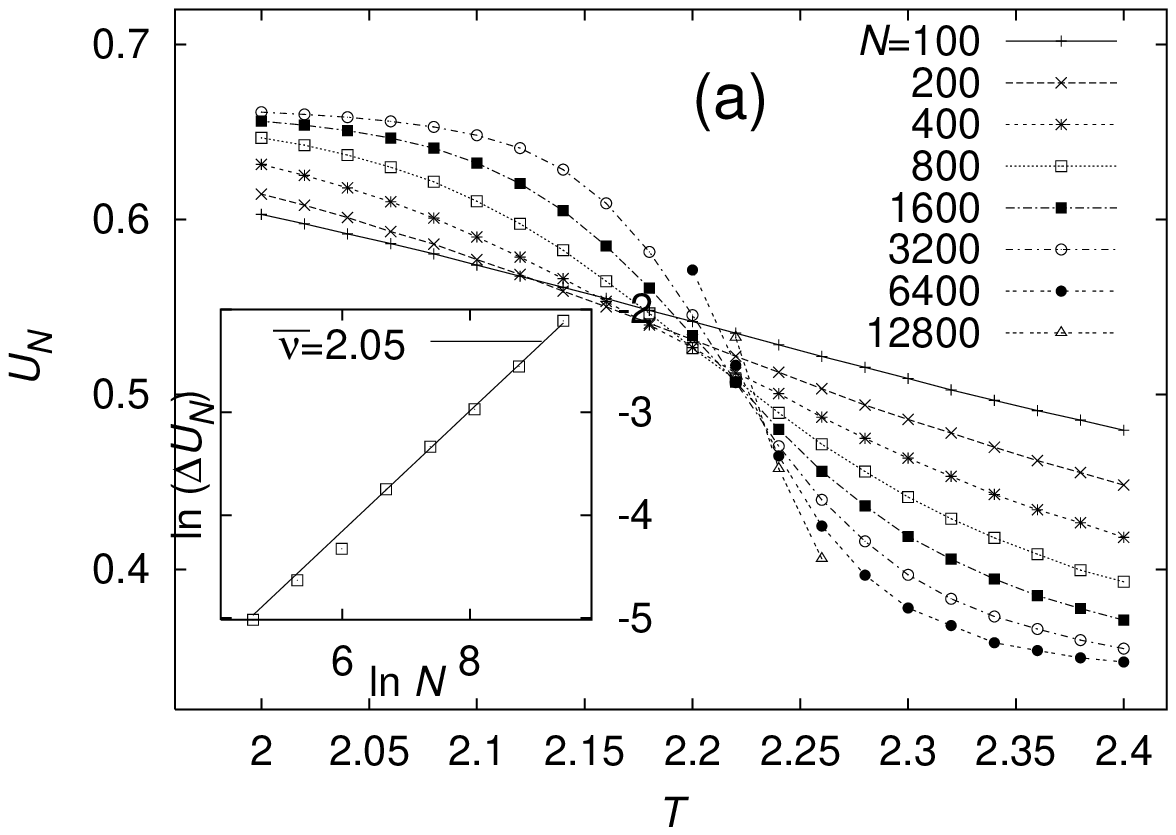}}}
\centering{\resizebox*{!}{6.0cm}{\includegraphics{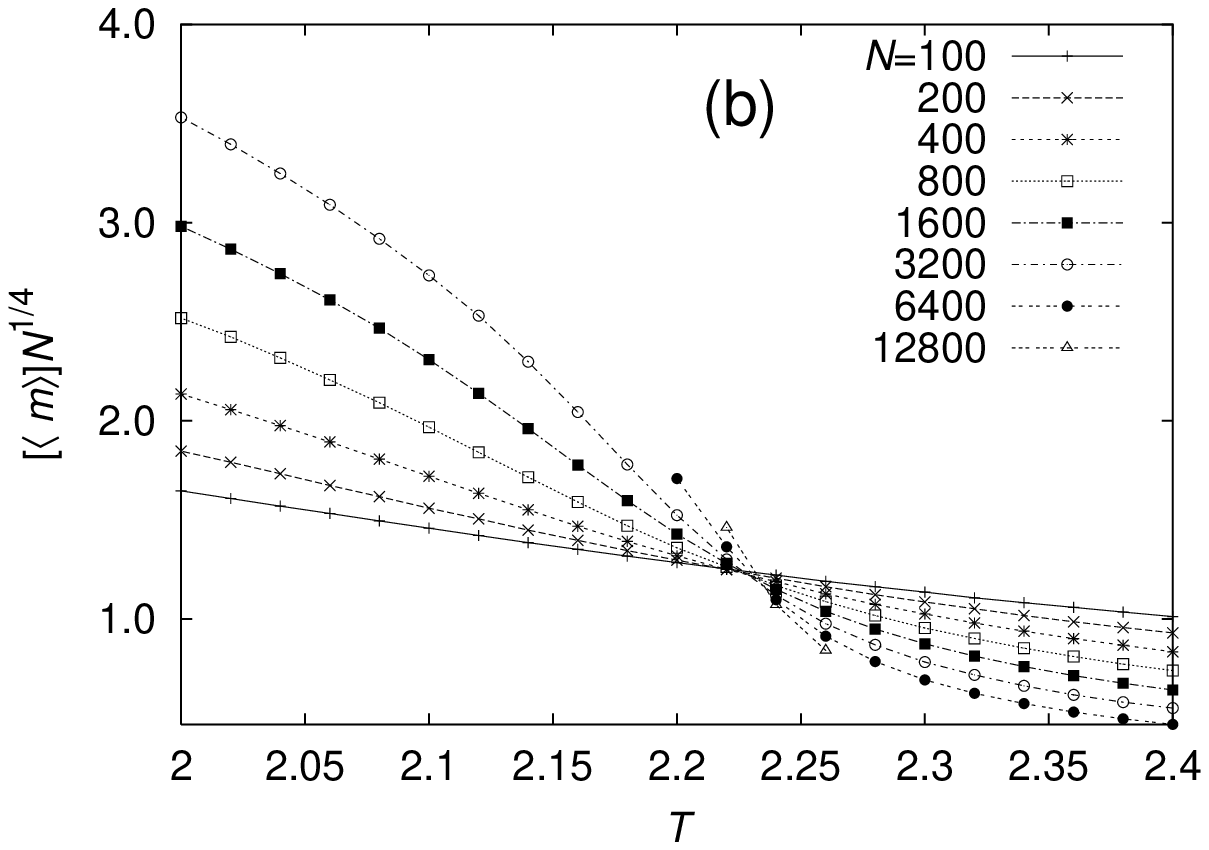}}}
\centering{\resizebox*{!}{6.0cm}{\includegraphics{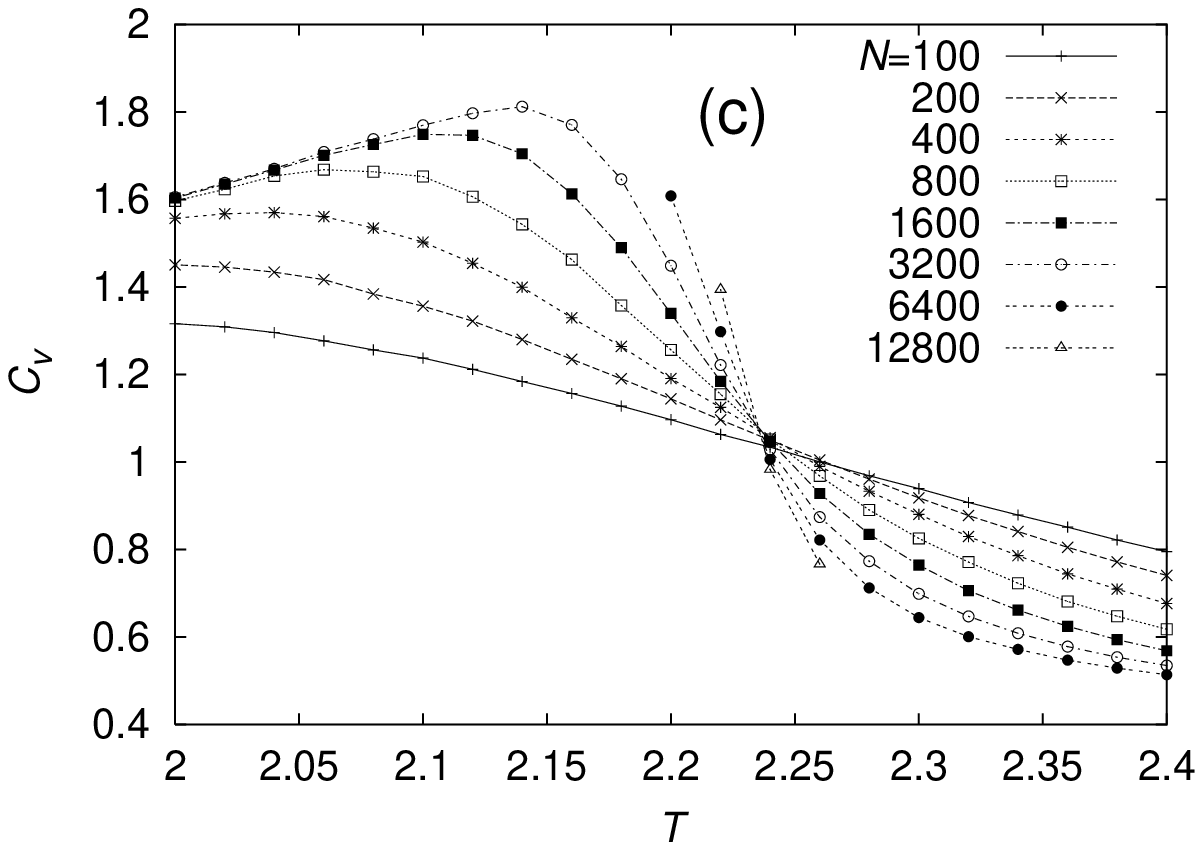}}}
\caption{The \xy model on the small-world network with $P=0.2$ and $k=3$
displays a mean-field phase transition at the finite temperature  $T_c$:
(a) Binder's cumulant $U_N$ has a unique crossing point
at $T_c\approx 2.235$ (in units of $J/k_B$) 
and the critical exponent $\bar\nu\approx 2.0$ 
is obtained from Eq.~(\ref{eq:Bexpand1}) (see the inset).  
(b) Finite-size scaling of the order parameter [see Eq.~(\ref{eq:mscale})]
again yields $\beta \approx 0.5$ and $\bar\nu \approx 2.0$ with 
$T_c \approx 2.235$.
(c) Specific heat $C_v$ (in units of $k_B$) 
also has a crossing point at $T_c \approx 2.235$,
as implied by Eq.~(\ref{eq:Cvscale}). 
}
\label{fig:p0.2}
\end{figure}

The small-world network, on which the \xy model is defined, is characterized by
the rewiring probability $P$ and the local interaction range $k$.
We focus only on the case $k=3$; 
however, we believe that qualitatively the same conclusion should hold 
for other values of $k \,(>1)$~\cite{foot_cluster}. 
From the MC simulations, we have confirmed that the network with $P=0$
does not exhibit long-range order at finite temperatures, in accordance with
the Mermin-Wagner theorem. 
In the opposite case with $P=1$, 
where all $kN \,(=3N)$ connections are long-ranged shortcuts, 
the system is found to undergo a well-defined finite-temperature transition.
We first present our MC results for $P=0.2$ in Fig.~\ref{fig:p0.2}.
The finite-size scaling of Binder's cumulant, the order parameter,
and the specific heat all reveals unanimously $T_c \approx 2.235$ in units
of $J/k_B$, confirming the presence of a finite-temperature phase transition
in the \xy model on the small-world network. 
In particular, the obtained critical exponents 
$\beta = 1/2$, $\bar\nu = 2$, and $\alpha = 0$ 
establish the mean-field nature of the transition. 
Also confirmed is the validity of our finite-size scaling forms,
with the additional length scale $\zeta$
reflected in the drift of the crossing points at small sizes.
As the size $N$ is increased, the finite-size effects associated with $\zeta$ 
reduce and the finite-size scaling forms based on the assumption
of only two scales $N$ and $\xi_V$ prevail, leading to well-defined
crossing points at larger sizes.

\begin{figure}
\centering{\resizebox*{!}{6.0cm}{\includegraphics{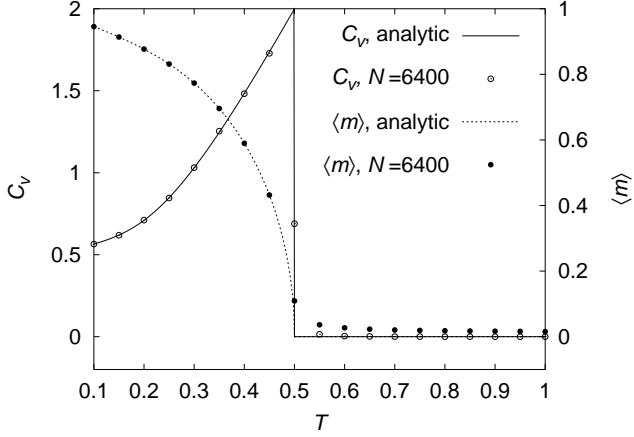}}}
\caption{Specific heat $C_v$ (in units of $k_B$) 
and the order parameter $\langle m\rangle$ 
of the globally-coupled \xy model.  Compared are $C_v$ and $\langle m\rangle$ 
from analytic calculation (full and dotted lines, respectively) 
and from Monte-Carlo simulations for the system size $N=6400$ 
(empty squares and filled circles). 
The phase transition is of the mean-field type characterized 
by $\langle m \rangle \sim (T_c - T)^\beta$ with $\beta = 1/2$
and the jump in $C_v$ at the transition ($T_c = 1/2$ in units of $JN/k_B$).
}
\label{fig:comp}
\end{figure}

For comparison, we also study the globally 
coupled \xy model, where $J_{ij} = J = {\cal O}(1/N)$ for all $i$ and $j$
and $T$ is measured in units of $NJ/k_B$. 
We use 
the saddle-point method accompanied
by the Hubbard-Stratonovich transformation~\cite{antoni}, 
and obtain various thermodynamic quantities such as the specific heat 
and the order parameter in Eq.~(\ref{eq:m}).
We also perform the MC simulation and compare the results
in Fig.~\ref{fig:comp}, which exhibits excellent correspondence between the two.  
As expected, the globally-coupled \xy model indeed possesses 
the mean-field transition, which is the same as that in the \xy model
on the small-world network.
It should be noted that the number of connected links 
(i.e., the number of nonzero elements in the interaction matrix $J_{ij}$) 
in the globally-coupled \xy model is given by $N(N{-}1)/2$, 
which is in sharp contrast with the number ${\cal O}(N)$ in the
small-world networks. 
This indicates that the \xy model can be made to have global phase
coherence (i.e., true long-range order) at finite temperatures 
by rewiring ${\cal O}(N)$ connections, 
with the total number of interactions still kept ${\cal O}(N)$
instead of ${\cal O}(N^2)$ as in the globally-coupled \xy model.

\begin{figure}
\centering{\resizebox*{!}{6.0cm}{\includegraphics{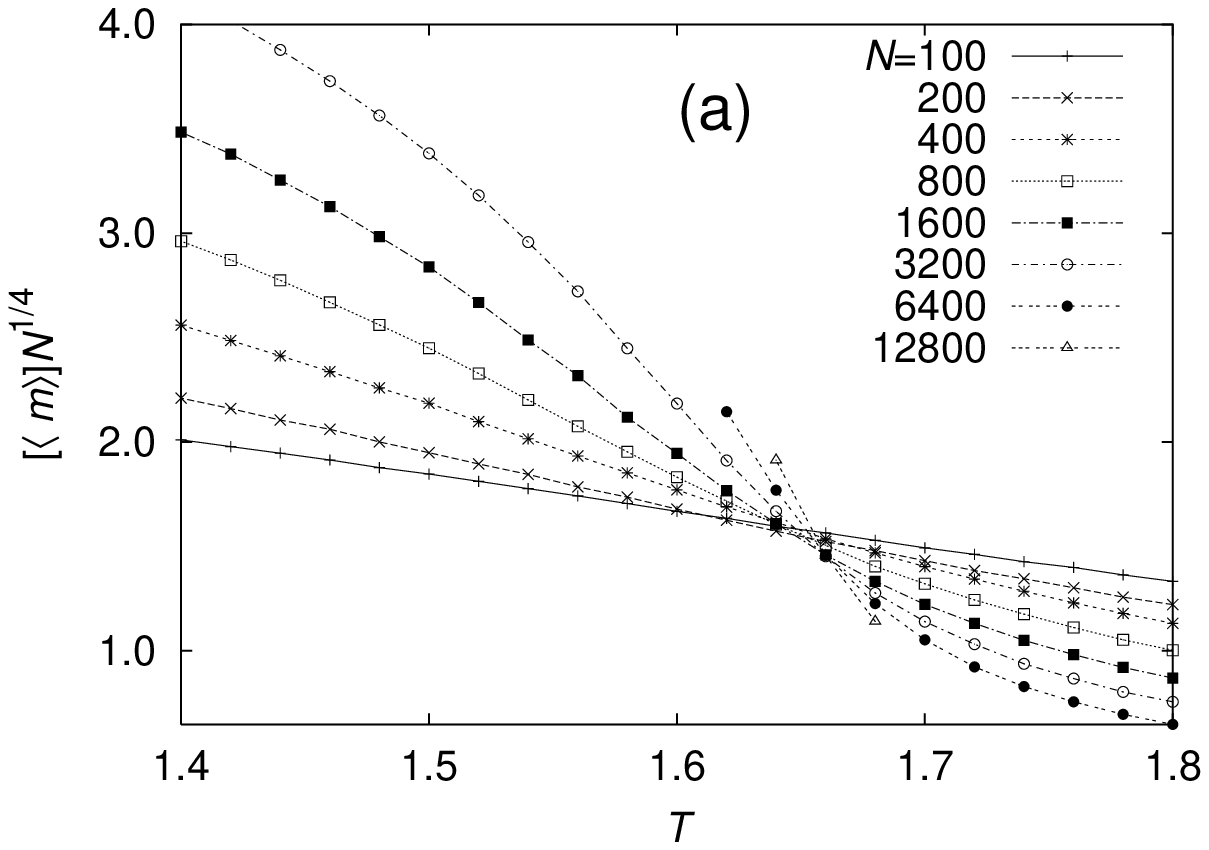}}}

\centering{\resizebox*{!}{6.0cm}{\includegraphics{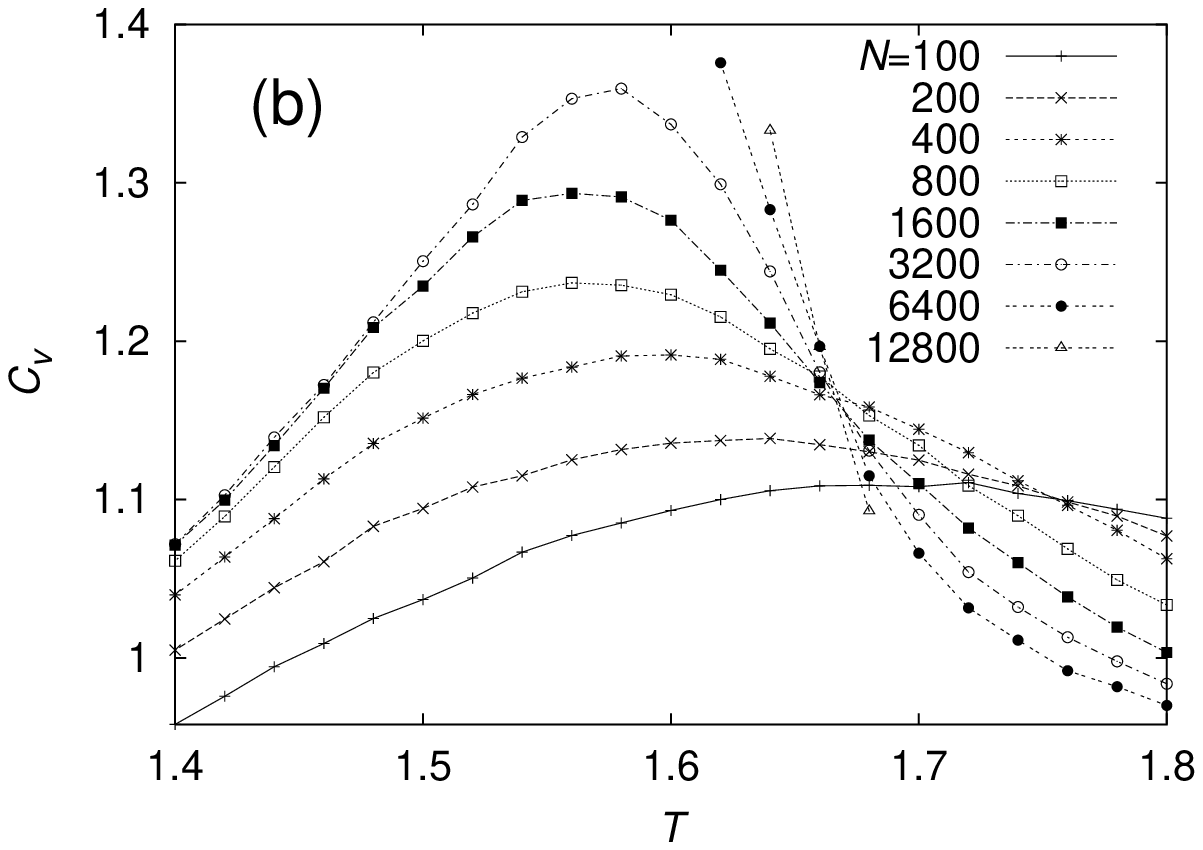}}}
\caption{\xy model on the small-world network for $P=0.05$.
From the crossing points in the scaling forms of (a) 
the order parameter $[\langle m \rangle]$
and (b) the specific heat $C_v$ (in units of $k_B$), $T_c \approx 1.66$ 
(in units of $J/k_B$) is obtained.
The drift of the crossing point at small sizes is larger 
in the specific heat (b) than in the order parameter (a).
}
\label{fig:p0.05}
\end{figure}

What is the minimum value of $P$ to have such long-rage order at 
finite temperatures? 
To answer this question, we have also performed MC simulations at 
various values of $P$.  As the rewiring
probability $P$ is reduced, the finite-size effects due to
the length scale $\zeta$ grow and become no more negligible,
making it necessary to perform simulations on larger systems. 
Since it is observed that
the finite-size effects are much more discernible in Binder's cumulant
and in the specific heat than in the order parameter, 
we estimate $T_c$ only from the crossing point in the plot 
of $[\langle m\rangle] N^{1/4}$ versus $T$.  
For $P \geq 0.03$, we have confirmed that as the system size grows, all the three 
methods to determine $T_c$ give the same value.
As an example, we show in Fig.~\ref{fig:p0.05} the determination of $T_c$ for $P=0.05$. 

\begin{figure}
\centering{\resizebox*{!}{6.0cm}{\includegraphics{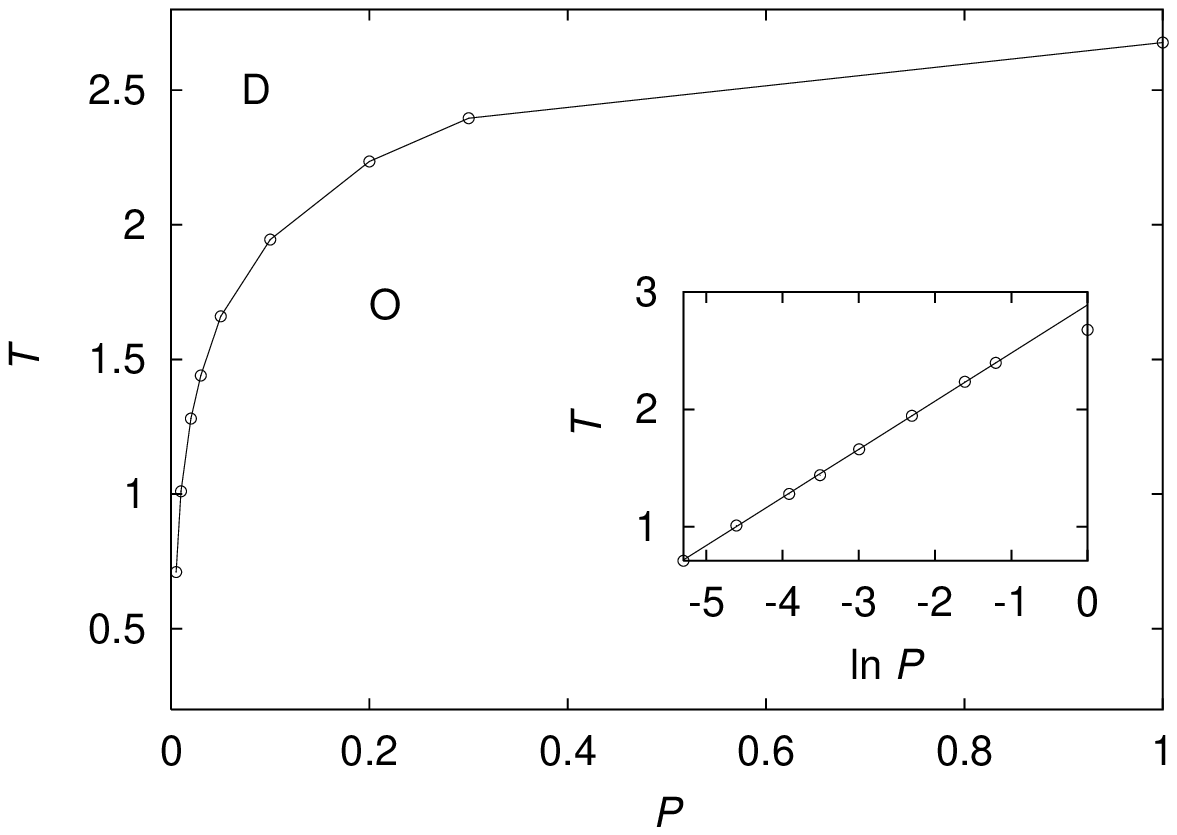}}}
\caption{The phase diagram of the \xy model on the small-world network with 
$k=3$, drawn on the plane of the rewiring probability $P$ and the temperature $T$ 
(in units of $J/k_B$). 
The data points have been obtained from the finite-size
scaling form of the order parameter in Eq.~(\ref{eq:mscale}) and the error
bars are smaller than the size of the symbol.  
The solid line, which is only a guide to eyes, gives the boundary
separating the disordered high-temperature phase (D) from the low-temperature
phase with the non-vanishing order parameter (O). 
Inset: The phase boundary is well described by the form
$T \approx 0.41\ln P + 2.89$ except for the point $P=1$.
}
\label{fig:phd}
\end{figure}

Finally, Fig.~\ref{fig:phd} summarizes the results and presents the phase diagram 
for the \xy model on small-world networks with $k=3$.
We observe that the phase boundary is very well described by the
logarithmic form: $T_c(P) \approx 0.41\ln P + 2.89$, as shown
in the inset of Fig.~\ref{fig:phd}, without an obvious reason.  
A naive extrapolation of this form predicts a very small value of 
the critical rewiring probability: $P_c\approx 0.001$. 
It is, however, likely that there takes place deviation 
from the logarithmic dependence at small $P$ 
and any nonzero rewiring probability ($P>0$) presumably supports
long-range order at sufficiently low but finite temperatures, 
resulting in $P_c = 0$. 
One may provide a simple argument in favor of this expectation: 
Since the spins separated within the correlation length are 
correlated with each other, the presence of shortcuts in such a way that
the typical distance $\zeta$ between the ends of shortcuts 
is smaller than the correlation length $\xi$ (of the 1D system) 
does not affect the system substantially, leaving the 1D nature intact. 
When $\zeta$ grows beyond the correlation length, on the other hand, 
the long-range interactions via shortcuts come into play, 
giving rise to the mean-field character. 
Accordingly, the crossover between the 1D behavior and the mean-field one is
given by the condition $\xi \approx \zeta$, which also 
describes the so-called small-world transition between the two geometrical regimes, 
the large-world regime and the small-world one~\cite{smtr}. 
Recognizing the low-temperature behavior of 
the correlation length $\xi \sim T^{-1}$~\cite{1DXY} 
in one dimension ($P = 0$) and the typical distance $\zeta \sim P^{-1}$, 
we thus expect the behavior of the transition temperature 
$T_c \sim P$ in the limit of small $P$~\cite{comm}. 
Here it is of interest to note the difference from the Ising model 
on small-world networks. 
The latter exhibits the behavior $T_c \sim -(\ln P)^{-1}$~\cite{ising}, 
which originates from the exponential temperature dependence in the 1D Ising model, 
$\xi \sim e^{a/T}$ with the constant $a$ proportional to the coupling strength. 

In conclusion, we have studied the \xy model on small-world
networks through the use
of the standard Monte-Carlo simulation method. 
At all nonzero values of the rewiring probability $P$ considered in this work, 
the finite-temperature transitions have been detected by Binder's cumulant, 
the specific heat, and the order parameter, 
together with appropriate finite-size scaling forms.
The phase transition has been shown to be described 
by the mean-field critical exponents, $\beta =1/2$, $\bar\nu = 2$,
and $\alpha = 0$, irrespective of the value of $P$.  
The existence of the additional length scale corresponding to the typical distance
between the ends of shortcuts
is manifested by the drift of crossing points at smaller sizes.
The question as to the value of $P_c$,
below which long-range order does not emerge at any finite temperature, 
still remains to be answered. 
The phase diagram in Fig.~\ref{fig:phd} 
provides a very small upper bound, $P_c \lesssim {\cal O}(10^{-3})$, 
which is likely to suggest $P_c = 0$.
In this regard, it is revealing that the small-world transition, 
where the characteristic path length $\ell$ undergoes a change of behavior, 
occurs at $P=0$ in the thermodynamic limit~\cite{smtr}. 
We thus suspect that the small-world transition and the order-disorder transition 
are intimately related, detailed investigation of which is left for further study.

This work was supported in part by the Swedish Natural Research Council
through Contract No.\ F 5102-659/2001 (B.J.K., P.H., and P.M.) and
by the Ministry of Education of Korea through the BK21 Program (H.H. and M.Y.C.).

\end{multicols}

\begin{thebibliography}{10}
\bibitem{foot:nubar} In $d$ dimensions, we have 
$\xi_V \sim \xi^d$ with the correlation length $\xi$, 
which diverges as $\xi \sim |T-T_c|^{-\nu}$.  
Accordingly, ${\bar \nu} = d\nu$ in a $d$-dimensional system.
\bibitem{foot_cluster} The WS model with $k=1$ is somewhat exceptional
since the clustering coefficient vanishes in this case.  
Some of the results in the present work may not hold for $k=1$.
\bibitem{comm} Although the crossover temperature provides an upper bound for 
the transition temperature, the two have been assumed to exhibit similar 
limiting behavior to zero as the rewiring probability is reduced to zero.
\end{thebibliography}
\end{document}